\documentclass[sigconf]{acmart}

\usepackage{graphicx}
\usepackage{float}
\usepackage{multirow}
\usepackage[export]{adjustbox} 
\usepackage{amsfonts} 

\usepackage{colortbl} 

\usepackage{bbm}
\definecolor{mygray}{gray}{0.6}
\definecolor{mygray-bg}{gray}{0.95}

\AtBeginDocument{%
  \providecommand\BibTeX{{%
    \normalfont B\kern-0.5em{\scshape i\kern-0.25em b}\kern-0.8em\TeX}}}

\copyrightyear{2023}
\acmYear{2023}
\setcopyright{rightsretained}
\acmConference[MM '23]{Proceedings of the 31st ACM International Conference on Multimedia}{October 29-November 3, 2023}{Ottawa, ON, Canada}
\acmBooktitle{Proceedings of the 31st ACM International Conference on Multimedia (MM '23), October 29-November 3, 2023, Ottawa, ON, Canada}
\acmDOI{10.1145/3581783.3612307}
\acmISBN{979-8-4007-0108-5/23/10}

\begin{document}

\title{DiffDance: Cascaded Human Motion Diffusion Model \\ for Dance Generation}

\author{Qiaosong Qi}
\authornote{Both authors contributed equally to this research.}
\email{qiqiaosong.qqs@alibaba-inc.com
}
\orcid{}
\affiliation{
  \institution{Alibaba Group}
  \city{Beijing}
  \country{China}
}

\author{Le Zhuo}
\authornotemark[1]
\email{zhuole1025@gmail.com}
\orcid{0000-0001-7895-091X}
\affiliation{
  \institution{Beihang University}
  \city{Beijing}
  \country{China}
}

\author{Aixi Zhang}
\authornote{Corresponding author.}
\email{aixi.zhax@alibaba-inc.com
}
\affiliation{
  \institution{Alibaba Group}
  \city{Beijing}
  \country{China}
}

\author{Yue Liao}
\email{liaoyue.ai@gmail.com}
\affiliation{
  \institution{Beihang University}
  \city{Beijing}
  \country{China}
}

\author{Fei Fang}
\email{mingyi.ff@alibaba-inc.com}
\affiliation{
  \institution{Alibaba Group}
  \city{Beijing}
  \country{China}
}

\author{Si Liu}
\email{liusi@buaa.edu.cn}
\affiliation{
  \institution{Beihang University}
  \city{Beijing}
  \country{China}
}

\author{Shuicheng Yan}
\email{shuicheng.yan@gmail.com}
\affiliation{
  \institution{BAAI}
  \institution{Skyworks}
  \city{}
  \country{}
}

\renewcommand{\shortauthors}{Qi and Zhuo, et al.}

\begin{abstract}
When hearing music, it is natural for people to dance to its rhythm. Automatic dance generation, however, is a challenging task due to the physical constraints of human motion and rhythmic alignment with target music. Conventional autoregressive methods introduce compounding errors during sampling and struggle to capture the long-term structure of dance sequences. To address these limitations, we present a novel cascaded motion diffusion model, DiffDance, designed for high-resolution, long-form dance generation. This model comprises a music-to-dance diffusion model and a sequence super-resolution diffusion model.
To bridge the gap between music and motion for conditional generation, DiffDance employs a pretrained audio representation learning model to extract music embeddings and further align its embedding space to motion via contrastive loss. During training our cascaded diffusion model, we also incorporate multiple geometric losses to constrain the model outputs to be physically plausible and add a dynamic loss weight that adaptively changes over diffusion timesteps to facilitate sample diversity.
Through comprehensive experiments performed on the benchmark dataset AIST++, we demonstrate that DiffDance is capable of generating realistic dance sequences that align effectively with the input music. These results are comparable to those achieved by state-of-the-art autoregressive methods.
\end{abstract}


\begin{teaserfigure}
    \captionsetup{type=figure}
    \begin{center}
    \includegraphics[width=1.0\linewidth]{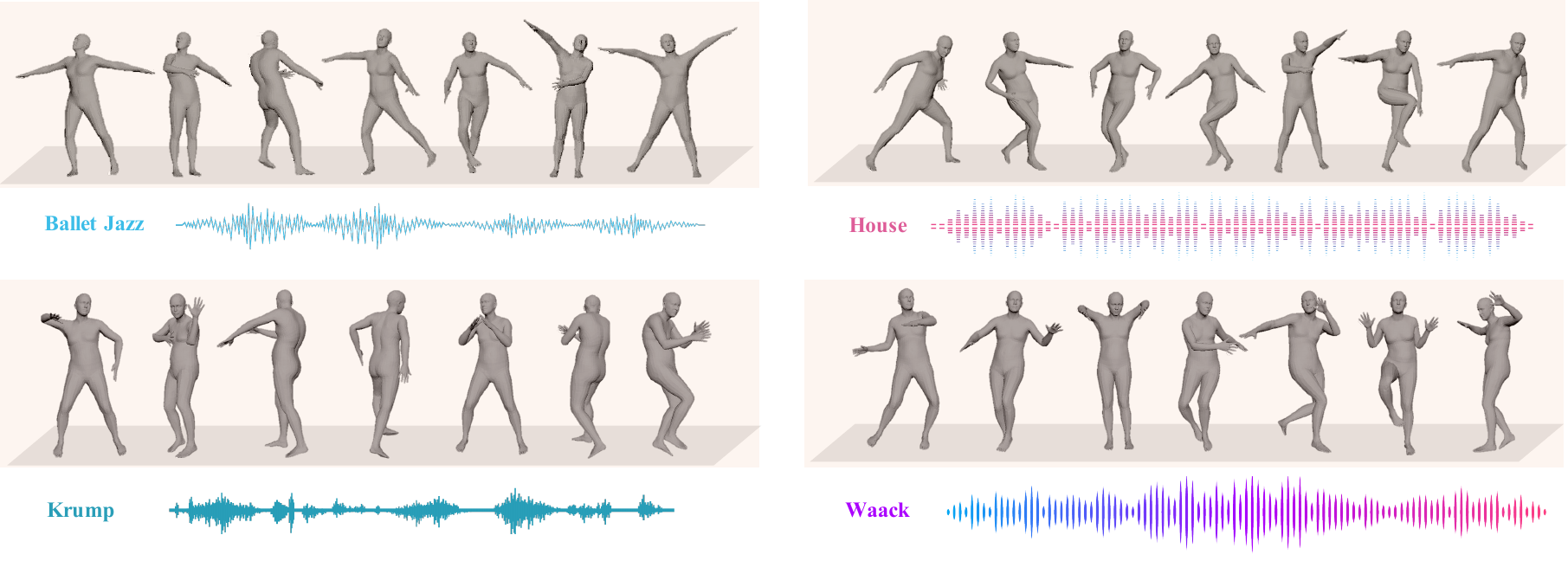}
    \captionof{figure}{Dance examples generated by our DiffDance conditioned on various genres of music. Our cascaded diffusion model framework, when conditioned on aligned music representations, is capable of producing high-resolution, long-form dance sequences that correspond well with the input music.}
    \label{fig:dance}
    \end{center}
\end{teaserfigure}

\maketitle


\section{Introduction}
\label{sec:intro}
\rightline{\emph{See the music, hear the dance.}}
\rightline{\emph{George Balanchine}}

Dance serves as a universal language that transcends not only cultural boundaries but also those between species. In recent times, dance videos have emerged as the most popular video category on social media platforms such as TikTok and YouTube. Dance is inherently intertwined with music, as individuals naturally move to the rhythm. However, creating a satisfactory dance is challenging, as it requires elegant movements that synchronize with the music on both a global style and local rhythm level, a feat that may take professional dancers years of practice. Consequently, the task of automatic dance generation from music has garnered significant interest from the deep learning community in recent years.

Existing works~\cite{LearningTG,DanceRL,aist++,bailando} primarily formulate music-to-dance generation as a sequence-to-sequence generation task that autoregressively generates dance sequences. However, these approaches trained with teacher forcing strategy are susceptible to compounding error introduced by autoregressive generation, which becomes problematic for generating long sequences. Besides, conventional methods rely on handcrafted spectrogram features as music conditions, including~\emph{MFCC},~\emph{onset strength},~\emph{constant-Q chromagram},~\emph{etc.} These features may lack a deep understanding of the music-dance relationship and could be suboptimal for music-to-dance generation.

Diffusion models, a newly developed class of non-autoregressive generative models, achieve impressive results in various tasks~\cite{dhariwal2021diffusion,kong2021diffwave,zeng2022lion}. For conditional synthesis, diffusion models also demonstrate a strong capacity to generate diverse and realistic samples~\cite{glide,saharia2022photorealistic}. Recent work~\cite{mdm} proposes Motion Diffusion Model (MDM), which achieves state-of-the-art results in text-to-motion and action-to-motion. However, we argue that directly applying existing motion diffusion models to dance generation is problematic since they are designed for motion with low temporal resolution and short sequence length. In contrast, dance sequences are usually much longer and more complex than general human motion, exhibiting global structures such as symmetrical repetitive movements. Therefore, these models struggle to model extremely long sequences and fail to produce realistic dance with long-term structure. 

In this work, we aim to generate high temporal resolution dance sequences aligned with input music leveraging diffusion models. To this end, we propose a novel cascaded human diffusion model framework named DiffDance. Specifically, DiffDance is a two-stage method that contains a music-to-dance diffusion model first generating low-resolution dance sequences and a sequence super-resolution model upscaling the low-resolution sequence by filling intermediate motion between input motions. To enable conditional generation, we use Wav2CLIP~\cite{wu2022wav2clip} to map input music into learned embeddings instead of conventional handcrafted features. Models in both stages are conditioned on the learned embeddings and use classifier-free guidance~\cite{classifier-free} to improve sample quality. Since the Wav2CLIP audio encoder shares an embedding space solely with images and text, we align its embedding space to motion by freezing the motion encoder in MotionCLIP~\cite{tevet2022motionclip} and fine-tuning our audio encoder using paired data in AIST++~\cite{aist++}. After fine-tuning, the audio encoder can produce latent representations aligned with motion semantics, thus further boosting music-to-dance performance. Lastly, we incorporate multiple geometric losses during training, as derived from existing motion generation literature~\cite{kocabas2020vibe,shi2020motionet,petrovich2021action}, introducing key joints position and rotation regularization losses to prevent unnatural artifacts such as foot sliding and instant rotation. We further add a dynamic loss weight to encourage model sampling at large timesteps and correcting unnatural motion at small timesteps. As illustrated in Fig.~\ref{fig:dance}, our DiffDance can generate diverse, realistic, and coherent dance sequences guided by various music inputs.

Our contributions can be summarized as follows:
\begin{itemize}
    \item We propose a cascaded motion diffusion model that generates long-form, high-resolution dance sequences;
    \item We align the CLIP embedding space of music and dance for better feature representation and demonstrate the effectiveness of classifier-free guidance in music-to-dance generation;
    \item We incorporate a variety of geometric losses and a dynamic loss weight schedule to produce realistic samples while maintaining diversity;
    \item Extensive experiments demonstrate our proposed DiffDance surpasses the state-of-the-art model in terms of dance quality and music-dance correspondence.
\end{itemize}

\section{RELATED WORK}
\label{sec:related}

\subsection{Music to Dance Generation}
Generating dance sequences from music, which aims to produce realistic choreographed movements aligned with input music, is a challenging task built on motion synthesis~\cite{butepage2017deep,hernandez2019human,aksan2019structured,duan2021automatic}. Early works~\cite{lee2018listen,Music2Dance,shlizerman2018audio} favored the generation of 2D dance sequences from music, predominantly due to the abundant data from online dance videos. Recently, AIST++~\cite{aist++}, a large-scale 3D motion dataset, greatly pushed the development of 3D dance generation. Various works explore this task leveraging different network architectures, such as LSTMs~\cite{yalta2019weakly}, Transformers~\cite{aist++,DanceRL,bailando}, and GANs~\cite{kim2022brand}. Among them, works with transformer architecture achieve state-of-the-art results, highlighting the superiority of Transformers in sequence modeling. For instance, FACT~\cite{aist++} introduced a full-attention cross-modal transformer that generates high-quality 3D dance sequences in an autoregressive manner. 
Bailando~\cite{bailando} designed a two-stage method consisting of pose VQ-VAE and motion GPT. The motion GPT is fine-tuned via actor-critic learning to realize temporal coherency. MNET~\cite{kim2022brand} proposed a conditional GAN framework including a Transformer Generator and Discriminator to produce dance motions conditioned on multiple music genres. In contrast, our method leverages a cascaded diffusion model framework to directly generate the whole dance sequences avoiding compounding errors introduced by autoregressive generation. 

It is also worth mentioning that existing methods largely overlook the aspect of music representation, typically directly using handcrafted music features extracted by Librosa~\cite{librosa} as conditional music features, such as~\emph{MFCC},~\emph{onset strength}, and the~\emph{constant-Q chromagram}. However, recent advancements in text-to-image and text-to-video generation~\cite{saharia2022photorealistic,ho2022imagen} have demonstrated that using conditional features extracted from large-scale representation learning models can markedly enhance cross-modal generation performance. Motivated by this finding, our model deviates from traditional handcrafted music features and instead employs the Wav2CLIP~\cite{wu2022wav2clip} audio encoder, a robust audio representation learning method, for music-driven dance generation, resulting in improved dance generation performance.

\subsection{Diffusion Models} 
Diffusion models~\cite{sohl2015deep,song2019generative,ho2020denoising} are a class of likelihood-based generative models that learn to recover samples from random noise via a denoising process. They have achieved great success in generating high-fidelity images~\cite{dhariwal2021diffusion}, and demonstrate strong abilities to generalize to other domains such as audio~\cite{huang2022fastdiff} and language~\cite{li2022diffusionlm}. For conditional generation, existing models often use classifier guidance~\cite{dhariwal2021diffusion} or classifier-free guidance~\cite{classifier-free} to improve sample synthesis quality. Recently, some seminal works~\cite{mdm,zhang2022motiondiffuse} adapt diffusion models to motion synthesis and achieve impressive results. Specifically, MDM~\cite{mdm} proposed a transformer-based diffusion model that leverages classifier-free guidance to solve text-to-motion synthesis. However, dance sequences are much more difficult for diffusion models to synthesize since they have longer sequence lengths and more complex movements. In the vision domain, cascaded diffusion models~\cite{ho2022cascaded,ho2022imagen} are effective methods that can generate high-resolution samples while keeping each sub-network relatively simple. Inspired by that, we solve music-to-dance generation via a cascaded motion diffusion model. Besides, we add multiple geometric losses during training in order to ensure the model output to be physically-plausible.

\section{METHOD}

\begin{figure*}[h]
    \centering
    \includegraphics[width=\linewidth]{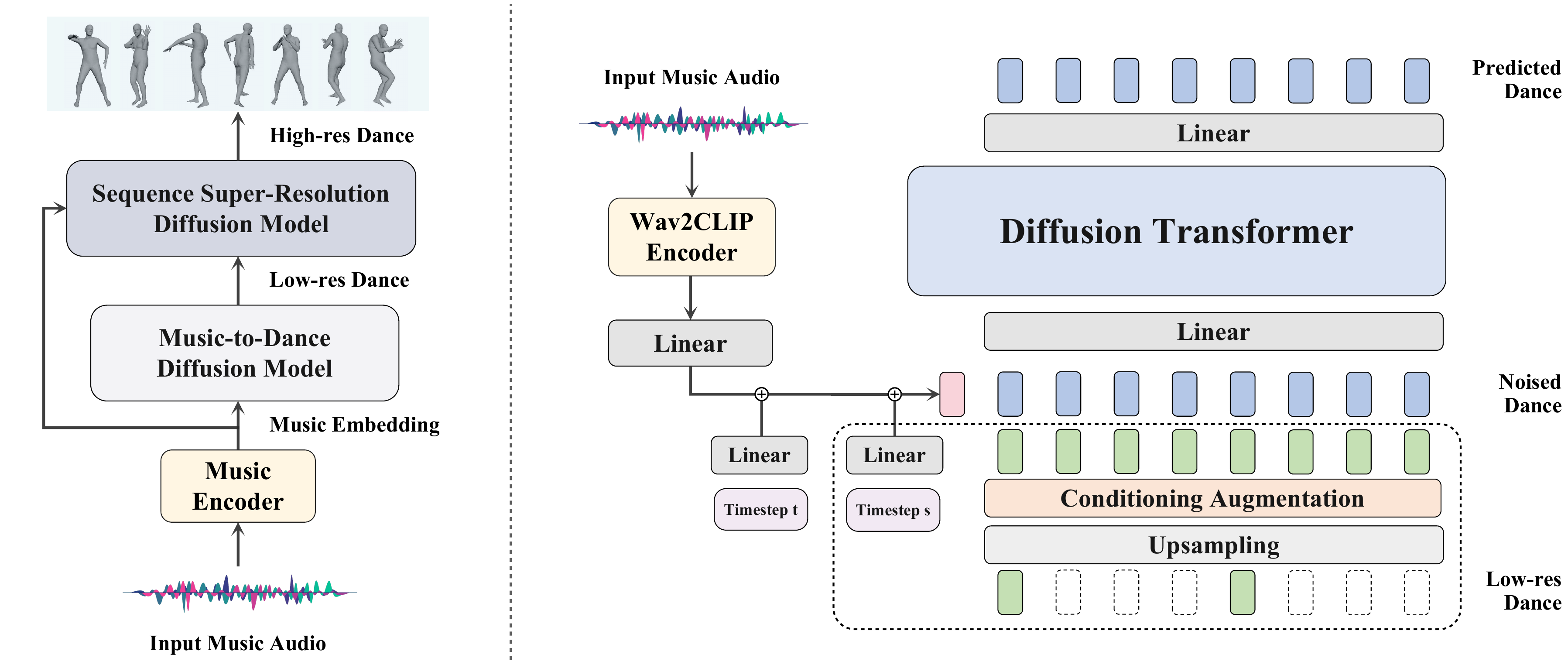}
    \captionof{figure}{(Left) Overall framework. DiffDance uses a frozen music encoder to extract music features. The music-to-dance diffusion model maps music representation into a low-resolution (15-FPS) dance sequence, while the sequence super-resolution diffusion model increases its temporal resolution (60-FPS). (Right) Model Details. Both models receive a noised dance $x_t$, diffusion timestep $t$, and music representation $c$. $c$ is extracted via Wav2CLIP encoder and then summed with $t$. For the sequence super-resolution model, it receives additional low-resolution dance $x_{low}$ and timestep $s$, as depicted in the dotted box. After conditional augmentation, low-resolution dance is channel-wise concatenated with noised dance.}
    \label{fig:framework}
    \vspace{-2mm}
\end{figure*}

\label{sec:method}
Our aim is to generate dance sequences $x^{1:L}=\{x^t\}^L_{t=1}$ with length $L$ given the music condition $c$. For 3D dance generation, the dance sequence is represented in $D$-dimensional features of $J$ body joints, resulting in a $x^t \in R^{J \times D}$ motion representation. In the following, we first briefly discuss the preliminary knowledge of diffusion models in Sec.~\ref{sec:preliminary}. Then, we introduce our proposed DiffDance, a cascaded motion diffusion model trained with multiple geometric losses in Sec.~\ref{sec:model}. In Sec.~\ref{sec:condition}, we finally discuss aligning music embedding space to motion and using classifier-free guidance for music conditional generation. 

\subsection{Preliminaries of Diffusion Models}
\label{sec:preliminary}
Diffusion models define a fixed Markovian process by gradually adding noise to sample data $x_0 \in q(x_0)$. The forward diffusion process is defined as:
\begin{align}
    q(x_{1:T}|x_0) &:= \sideset{}{}\prod_{t=1}^T q(x_t|x_{t-1}), \\
    q(x_{t}|x_{t-1}) &:= \mathcal{N}(x_t; \sqrt{1-\beta_t}x_{t-1}, \beta_t\mathbf{I})\label{q_xt},
\end{align}
where $t \in [1,T]$, $\beta_t$ is a pre-defined variance schedule that controls the rate of noise injection, and $q(x_{t}|x_{t-1})$ is a Gaussian transition kernel. After $T$ steps, the amount of noise becomes sufficiently large, and the Markovian chain approximately converges to a standard Gaussian distribution $\mathcal{N}(\mathbf{0}, \mathbf{I})$.

In order to convert noise back to sample for generation, diffusion models learn a reverse process via:
\begin{align}
    p_{\theta}(x_{0:T}) &:= p(x_T) \sideset{}{}\prod_{t=1}^T p_{\theta}(x_{t-1}|x_t),\label{p_x0} \\
    p_{\theta}(x_{t-1}|x_t) &:= \mathcal{N}(x_{t-1}; \mu_{\theta}(x_t,t), \Sigma_{\theta}(x_t, t)),\label{p_theta}
\end{align}
where $\theta$ is a parameterized neural network to predict the mean of Gaussian. In practice, we simply set $\Sigma_{\theta}(x_t, t) = \beta_t^2\textbf{I}$ following~\cite{ho2020denoising}.

Diffusion models are trained by minimizing the variational upper bound on the negative log-likelihood of data. Typically, we use a simplified version with $L_2$ loss~\cite{ho2020denoising}:
 \begin{equation}
    \mathcal{L}_{simple} = \mathbb{E}_{x_0 \sim q(x_0), t \sim [1,T]}\left[\left|\left|\hat{\epsilon}_{\theta}(x_t, t) - \epsilon  \right|\right|_2^2\right].
 \label{loss}
 \end{equation}
 
Once the diffusion model is trained, we can generate a new sample $x$ by iteratively running the reverse diffusion process from timestep $T$ to $0$.

\subsection{Cascaded Motion Diffusion Model}
\label{sec:model}
In this section, we formulate our DiffDance, a cascaded motion diffusion model for music-to-dance generation. Fig.~\ref{fig:framework} summarizes the cascaded pipeline of our DiffDance.

\noindent\textbf{Framework.} 
DiffDance consists of a Music-to-Dance (M2D) diffusion model and a Sequence Super-Resolution (SSR) diffusion model. Specifically, the M2D model is similar to MDM~\cite{mdm}, a transformer-based diffusion model as illustrated in Fig.~\ref{fig:framework}. Different from conventional diffusion models that predict $\epsilon_t$, our model $f^{low}_{\theta}(x_t, t, c)$ directly predicts the original data point $x_0$, given the noised data $x_t$, timestep $t$, and music representation $c$. Both the timestep $t$ and music representation $c$ are projected into the dimension of transformer. Next, we sum these embeddings together and concatenate it with noised input $x_t$ to guide the generation of our model. 

The SSR model $f^{high}_{\theta}(x_t, t, c, x_{low}, s)$ aims to increase the temporal resolution of the output of our base M2D model. It shares the same architecture with the M2D model except that there is an additional low-resolution dance input $x_{low}$ and an additional timestep $s$ for conditioning augmentation as shown in the dotted box of Fig.~\ref{fig:framework}. The SSR model first upsamples the low-resolution sequence via linearly interpolated motion frames and then channel-wise concatenates the upsampled input to the noised data $x_t$. We employ conditioning augmentation~\cite{ho2022cascaded} to our SSR model. Conditioning augmentation has been proven an effective strategy to significantly improve the sample quality and model robustness in cascaded generation pipelines~\cite{ho2022cascaded,saharia2022photorealistic}. In practice, we add Gaussian noise corresponding to a random diffusion timestep $s$ to corrupt input motion during training. We also add $s$ to the original diffusion timestep $t$ as conditional information. At inference time, we sweep over all possible values of $s$ and fix it that yields the best quality. Adding some noise to the inputs can eliminate unnatural artifacts in generated low-resolution dance sequences, thus bridging the gap between the ground-truth distribution during training and the model output distribution at inference time. 

\noindent\textbf{Training Objectives.}
Generating physically-plausible motion sequences is challenging using the original denoising objective with Equation~\ref{loss}. MDM adds a set of geometric losses~\cite{petrovich2021action,shi2020motionet} to regularize the training process. These losses, denoted as $L_{pos}^{a}$, regularize the positions and position velocities of all joints equally. 
However, we observe undesirable motions, such as instant movements and rotations generated by our vanilla model trained with $L_{pos}^{a}$, which are unreasonable for dance choreography. 
To produce fluent and natural dance sequences, we further add $L_{pos}^{k}$ to regularize key joints such as `hand' and `foot'. This regularization is also from both positions and position velocities perspectives formulated as:
 \begin{equation}
\mathcal{L}_{pos}^{k} = \sum_{j \in [f,h]} \| \mathbf{p}_j - \hat{\mathbf{p}}_j\|_{2}^{2} + \| \mathbf{v}^p_j - \hat{\mathbf{v}}^p_j\|_{2}^{2},
\label{key_pos}
 \end{equation}
where $f$ and $h$ denote key joints of foot and hand joints respectively, and $\mathcal{L}_{pos}^{k}$ constrains the difference between key joint positions $\hat{\mathbf{p}}_j$ and ground truth $\mathbf{p}_j$ as well as position velocities $\hat{\mathbf{v}}^p_j$ and ground truth $\mathbf{v}^p_j$. To account for the high complexity of motion sequences, we computed joint linear velocities with respect to the `root' node’s relative velocities. Therefore, our positions' loss function excludes the `root' joint’s linear velocity. 

Besides, we propose to regularize the rotations for key joints explicitly. Note that the `root' joint is also regularized as a key joint for barycentric motion consistency. The rotation loss is formulated as:
 \begin{equation}
 \mathcal{L}_{rot}^{k} = \sum_{j \in [f,h,r]} \| \mathbf{r}_j - \hat{\mathbf{r}}_j\|_{2}^{2} + \| \mathbf{v}^r_j - \hat{\mathbf{v}}^r_j\|_{2}^{2},
 \label{key_rot}
  \end{equation}
where $r$ denotes key joints of `root', and $\mathcal{L}_{rot}^{k}$ constrains the difference between key joint rotations $\mathbf{r}_j$ and ground truth $\hat{\mathbf{r}}_j$ as well as rotation velocities $\mathbf{v}^r_j$ and ground truth $\hat{\mathbf{v}}^r_j$.

We also find it unsuitable to apply these losses uniformly across all diffusion timesteps. As described in the analysis in~\cite{deja2022analyzing}, the backward diffusion process can be roughly divided into a generation stage and a denoising stage, corresponding to large and small timesteps, respectively. When $t$ approaches the total timestep $T$, the noise becomes sufficiently large, corrupting the input dance into a noised version that has almost lost all its geometric information. Intuitively, implementing geometric losses at the generation stage will not impose physical constraints on samples but may instead impair sample quality and diversity. To address this, we introduce a simple dynamic loss decay weight $\lambda_t=1-\alpha\frac{t}{T}$, which linearly decreases as the diffusion timestep $t$ increases. The overall training loss can be expressed as:
 \begin{equation}
    \mathcal{L} = \mathcal{L}_{simple} + \lambda_t(\lambda_1 \mathcal{L}_{pos}^{a} + \lambda_2 \mathcal{L}_{pos}^{k} + \lambda_3 \mathcal{L}_{rot}^{k}),
 \end{equation}
where $\lambda_1$, $\lambda_2$ and $\lambda_3$ are the hyper-parameter loss weights for the all joints position loss $\mathcal{L}_{pos}^{a}$, key joints position loss $\mathcal{L}_{pos}^{k}$ and key joints rotation loss $\mathcal{L}_{rot}^{k}$ respectively.

\subsection{Conditional Generation}
\label{sec:condition}
Given input music, our objective is to extract an effective music representation containing rich semantic information, such as the style and rhythm of the music. Subsequently, we utilize this representation as conditions to generate dance sequences aligned with the input music.

\begin{figure}[t]
\begin{center}
    \includegraphics[width=\linewidth]{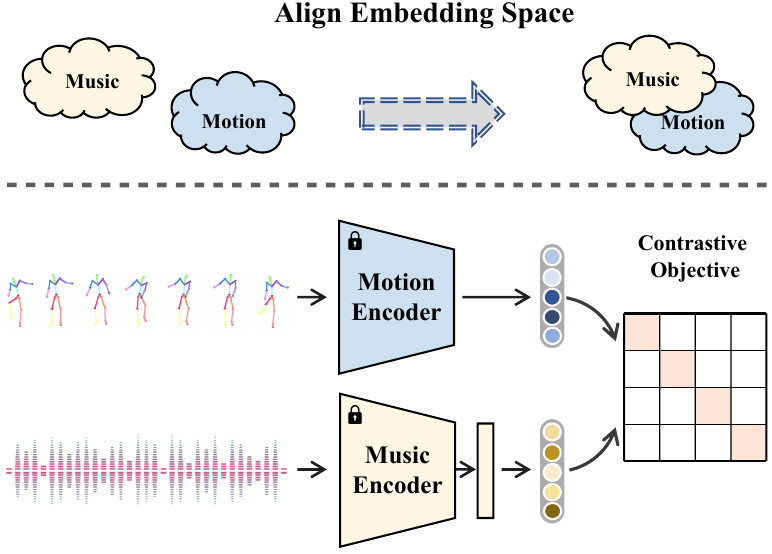}
    \caption{Alignment of music and motion. We introduce adapter layers to the music encoder and employ a contrastive loss to align the embedding spaces of music and motion.}
    \label{fig:align}
    \vspace{-3mm}
\end{center}
\end{figure}

\noindent\textbf{Music Representations.}
Previous approaches have not placed significant emphasis on music representations for conditional generation. These methods primarily employ handcrafted music features,~\emph{e.g.},~\emph{onset strength} as rhythmic features, and~\emph{constant-Q chromagram} as chroma features. One distinct drawback of these features is that they lack high-level semantics critical to cross-modality generation. CLIP~\cite{clip}, a large-scale visual-textual embedding model, has demonstrated its efficacy for text-guided generation works~\cite{saharia2022photorealistic,ho2022imagen}. Likewise, we propose to use the audio encoder of Wav2CLIP~\cite{wu2022wav2clip}, which encodes an audio clip into a $512$-dimensional vector that shares an embedding space with text and image in CLIP. However, there still exists a significant domain gap between extracted music representations and dance motions since the Wav2CLIP audio encoder is solely trained on general audio-visual datasets. To better align the embedding space of music audio with dance motions, we fine-tune our audio encoder by adding multi-layer perceptrons as adapter layers and map its output to a motion encoder, as illustrated in Fig.~\ref{fig:align}.

Specifically, we use the motion encoder in~\cite{tevet2022motionclip}, which also extracts a $512$-dimensional embedding aligned with CLIP joint representation. To mitigate modal collapse, we freeze the weights of both the motion encoder and music encoder during fine-tuning and only train the adapter layers with InfoNCE loss~\cite{infonce}. The music-to-dance contrastive loss for the $i$-th pair between music clip $m_i$ and dance sequence $d_i$ is formulated as:
\begin{equation}
\small
    \mathcal{L}^{m\rightarrow{}d}_i = - \log\frac{\exp\left[s(m_{i},d_{i})/\tau\right]}{{\sum_{j=1}^N\exp{\left[s(m_{i},d_{j})/\tau\right]}}},
    \label{eq:infonce}
\end{equation}
where $s(m_{i},d_{i})$ represents the cosine similarity, and $\tau$ is a learnable temperature parameter.

\noindent\textbf{Classifier-free Guidance.}
After learning the music-dance joint embedding space, we extract music representations and leverage classifier-free guidance~\cite{classifier-free} to improve sample quality. In practice, we jointly train a single diffusion model on both conditional and unconditional objectives by randomly dropping the music condition $c$. During sampling, we can improve sample quality by adjusting the $x_0$ prediction using:
\begin{align}
f_{\theta}(x_t, t, c) = w f_{\theta}(x_t, t, c) + (1 - w) f_{\theta}(x_t, t, \emptyset),
\end{align}
where $f_{\theta}(x_t, t, c)$ and $f_{\theta}(x_t, t, \emptyset)$ correspond to the conditional and unconditional model respectively, and $w$ is the guidance strength which is typically set greater than $1$ to enable classifier-free guidance. We apply classifier-free guidance for both models in the two-stage pipeline.

\section{EXPERIMENTS}


\subsection{Dataset}
AIST++~\cite{aist++} is a large-scale publicly available 3D human dance dataset containing $1363$ 3D dance sequences and music pairs. There are $980$ training sets, $40$ test sets, and $343$ candidate sets in AIST++. Note that the candidate sets are not recommended by AIST++ for training or evaluation. In terms of data format, dance motion is represented as $60$-FPS 3D pose sequences in SMPL format~\cite{smpl}. We conduct all the experiments on the AIST++ dataset and follow the experimental setting in~\cite{bailando}.

\subsection{Implementation Details}

\begin{table*}
    \centering
    \begin{tabular}{lcccccc}
        \hline
        \multirow{2}{*}{Method} &\multicolumn{2}{c}{Motion Quality} & \multicolumn{2}{c}{Motion Diversity} & \multirow{2}{*}{Beat Align Score$\uparrow$}  & User Study \\
         \cline{2-3} \cline{4-5} \cline{7-7}
         & FID$_k$$\downarrow$ & FID$_g$$\downarrow$ & Div$_k$$\uparrow$ & Div$_g$$\uparrow$ & & Our Method Wins \\
        \hline
        Ground Truth     & 17.10     & 10.60 & 8.19     & 7.45     & 0.2374 & 44.8\%  \\
        \hline
        Li~\emph{et al.}~\cite{ferreira2021learning}     & 86.43     & 43.46 & 6.85$^*$     & 3.32     & 0.1607 & 98.0\% \\
        DanceNet~\cite{Music2Dance}             & 69.18     & 25.49 & 2.86     & 2.85     & 0.1430 & 92.0\% \\
        DanceRevolution~\cite{DanceRL}      & 73.42     & 25.92 & 3.52     & 4.87     & 0.1950 & 90.8\% \\
        FACT~\cite{aist++}                 & 35.35     & 22.11 & 5.94     & \underline{6.18}     & 0.2209 & 96.8\% \\
        Bailando~\cite{bailando}             & \underline{28.16}     & \textbf{9.62}  & \textbf{7.83}     & \textbf{6.34}     & \underline{0.2332} & 74.4\% \\
        \cellcolor{mygray-bg}DiffDance (Ours)  & \cellcolor{mygray-bg}\textbf{24.09}  & \cellcolor{mygray-bg}\underline{20.68} & \cellcolor{mygray-bg}\underline{6.02}   & \cellcolor{mygray-bg}2.89   & \cellcolor{mygray-bg}\textbf{0.2418} & \cellcolor{mygray-bg} - \\
        \hline
    \end{tabular}
    \caption{\textbf{Baselines comparison on AIST++.} Best values are in bold, and runner-up values are underlined. Quantitatively, our model achieves state-of-the-art performance on FID$_k$ and Beat Align Score. Qualitatively, our model generates more realistic dance sequences and outperforms baseline approaches in the user study. $\downarrow$ indicates lower is better, and $\uparrow$ indicates higher is better. $*$Note that Li~\emph{et al.} generates discontinuous and jittery motion, leading to abnormally high Div$_k$, which is also reported in~\protect\cite{aist++,bailando}.}
    \label{tab:tabel1}
    \vspace{-2mm}
\end{table*}

\noindent\textbf{Alignment Setting.} For fine-tuning the Wav2CLIP adapter which consists of $2$ MLP layers with $512$ hidden size, we use AdamW~\cite{loshchilov2017decoupled} with learning-rate $1e^{-5}$ and train $100$ epochs with batch size $64$. Music is loaded by Librosa~\cite{librosa} and split into multiple clips of $6$ seconds. Dance sequences are represented in rotation 6d format~\cite{zhou2019continuity} and also split into $6$-second clips correspondingly. Similar to the experimental setting in MotionCLIP~\cite{tevet2022motionclip}, we down-sample the frame rate of dance clips from $60$-FPS to $30$-FPS.

\noindent\textbf{Cascaded Diffusion Model Setting.} We train our cascaded diffusion model for $500$ epochs using AdamW with learning-rate $1e^{-4}$. Music is loaded by Librosa and split into $20$-second clips, and the dance sequences are split correspondingly with rotation 6d representation. Music is mapped to $512$-dim vectors via frozen Wav2CLIP-adapter. For classifier-free guidance, we randomly mask $10\%$ music condition $c$ at each training step. The diffusion transformer has $12$ layers with $768$-dim hidden size and $6$ heads. The dropout ratio is set to $0.1$. All the geometric losses weights $\lambda_1$, $\lambda_2$, and $\lambda_3$ are set to $1.0$, and the decay coefficient $\alpha$ for $\lambda_t$ is set to $0.1$. For the base M2D model at the first stage, we set batch size to $32$. We use dance sequences of $20$ seconds to learn the long-term dance semantics and down-sample FPS from $60$ to $15$. For the SSR model at the second stage, we set batch size $8$, and keep FPS for high-resolution dance sequences to the default value of $60$. The whole cascaded framework trains on $4$ Tesla V100 GPUs in $24$ hours. 

In the inference stage, we generate a $20$-second ($1200$ frames) dance sequence, guided by a $2$-second ($120$ frames) seed dance sequence, and set the classifier-free guidance weight $w$ to $2.5$. We set the inference diffusion timestep $T$ to $100$, as we observe no significant difference between samples generated using the original timesteps and the reduced timesteps of $100$s.

\subsection{Evaluation Metrics}
We follow previous works~\cite{bailando,aist++} to quantitatively evaluate the generated samples in terms of dance quality, dance diversity, and music beat alignment. To evaluate dance quality, we first extract kinetic features~\cite{fid-k} and geometric features~\cite{fid-g} of ground truth and generated samples. Then we calculate Frechet Inception Distance (FID)~\cite{heusel2017gans} score, including FID$_k$ based on kinetic features and FID$_g$ based on geometric features. For dance diversity, we calculate the average feature distance of kinetic features and geometric features following~\cite{aist++}, denoted as Div$_k$ and Div$_g$, respectively. As for dance-music consistency, we use Beat Alignment Score (BAS) introduced in~\cite{aist++}, which calculates the average distance between music beat and its nearest dance beat:
\begin{equation} \label{eq:bas}
    \text{BAS} = \frac{1}{m} \sum_{i=1}^{m} \exp \bigg(-\frac{\min_{\forall b_{j}^{d} \in B^{d}} \parallel b_{i}^{d} - b_{j}^{m} \parallel^2}{2{\sigma}^2}\bigg), 
\end{equation}
where $B^d=\{b^d_i\}$ is the dance beats defined as the local minima of the kinetic velocity, $B^m=\{b^m_i\}$ is the music beats extracted using Librosa~\cite{librosa} toolbox, and $\sigma$ is a normalized parameter which is set to $3$ in all the experiments.

\begin{figure}[t]
\begin{center}
    \includegraphics[width=\linewidth]{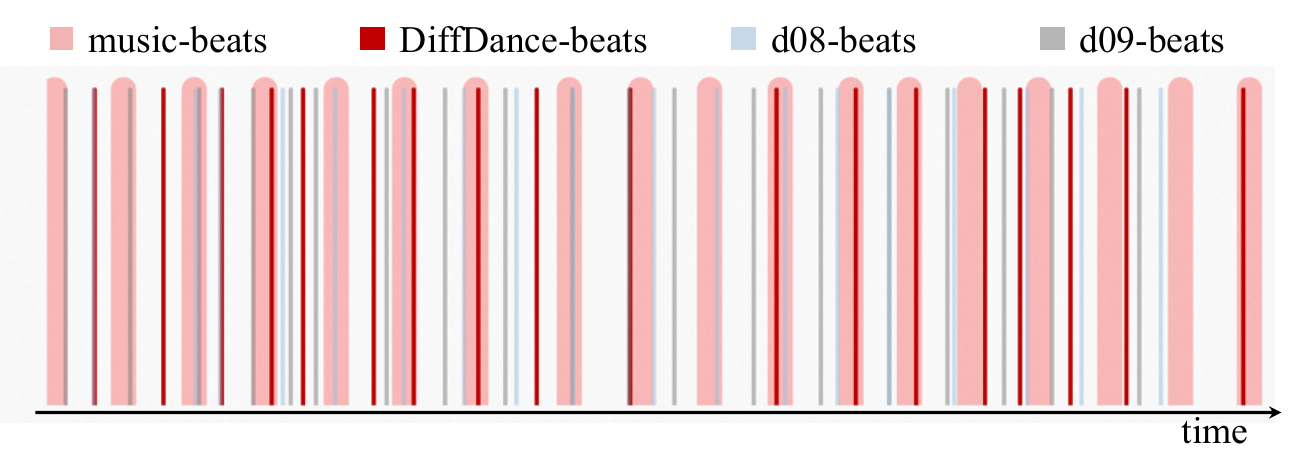}
    \caption{Beats alignment visualization. We present a visualization of music beats, kinematic beats of generated dances, and two ground truth dances. The distances between generated beats and music beats are smaller, indicating better rhythmic alignment.
    }
    \label{fig:beats_align}
    \vspace{-3mm}
\end{center}
\end{figure}

\subsection{Comparison with Baselines}
We mainly compare our proposed DiffDance with several existing methods, including Li~\emph{et al.}~\cite{LearningTG}, DanceNet~\cite{Music2Dance}, DanceRevolution~\cite{DanceRL}, FACT~\cite{aist++} and current state-of-the-art method Bailando~\cite{bailando}. Following~\cite{bailando}, we generate $40$ dance sequences for $20$-second music clips in AIST++ test set and calculate the evaluation metrics. According to the comparisons shown in Table~\ref{tab:tabel1}, our proposed DiffDance demonstrates comparable generative ability with state-of-the-art approaches, achieving $2$ best and $2$ runner-up in all $5$ objective evaluation metrics. Moreover, more users prefer our generated dance sequences compared to other methods in the user study. 

\noindent\textbf{Motion Quality.} As shown in Table~\ref{tab:tabel1}, our DiffDance achieves FID$_k$ score of $24.09$ and FID$_g$ score of $20.68$. Compared with existing methods, our model achieves the best FID$_k$, which outperforms $14.45\%$ with a margin of $4.07$ than the state-of-the-art method Bailando. This indicates that the kinetic features of generated samples, which guarantee dance characteristics, including motion velocities and energies, are much closer to those of ground truth dance distribution. As for the FID$_g$ score, which reflects the quality of choreography units, we achieve the second-best performance, which is $6.47\%$ better than FACT with a margin of $1.43$. 

We investigated the reasons that the FID$_g$ score of Bailando is much better than other methods, including ours. Firstly, Bailando adopts a Choreographic Memory Codebook to record and quantize dancing units from the $980$ dances of the AIST++ training set, which remembers the inherent dancing units of AIST++ to a certain extent. Secondly, all $1363$ dances in AIST++ dataset are used during evaluation (the train/test split is based on music-dance pairs). As a result, more than $70\%$ ground truth dancing units of the evaluation set have been memorized and quantized during training. The above two aspects will result in an overestimation of the FID$_g$ score, which is affected by the distance of geometric features between generated dance sequences and ground truth. As reported in Bailando, the quantization of dancing positions is essential, which helps FID$_g$ score improve from $147.28$ to $9.2$ with a considerable margin. 
Therefore, we argue that Bailando overestimates FID$_g$ evaluation. Our DiffDance is comparatively more general, as our model does not directly memorize dance units in the dataset, and the FID$_g$ score of DiffDance still outperforms other methods except Bailando.

\noindent\textbf{Motion Diversity.} Motion diversity is represented as the average Euclidean distance of generated dances in the kinetic and geometric feature spaces. Table~\ref{tab:tabel1} shows that our DiffDance achieves Div$_k$ of $6.02$ and Div$_g$ of $2.89$. The diversity of geometric features underperforms several previous methods partly due to the introduction of multiple regularization losses, which might limit the solution space of generated dance even with a dynamic loss weight. Besides, the guidance strength $w$ used in classifier-free guidance also has a trade-off between sample quality and diversity. For $w > 1$, this over-emphasizes the importance of condition $c$ during sampling, which might lead to higher quality but less diverse samples.

\begin{table}
    \centering
    \begin{tabular}{lccc}
        \hline
        Method  & FID$_k$$\downarrow$ & Div$_k$$\uparrow$ & BAS$\uparrow$ \\
        \hline
        Ground Truth     & 17.10     & 8.19  & 0.2374  \\
        DiffDance   & \textbf{24.09}         & \textbf{6.02}     & \textbf{0.2418} \\
        \quad w/o two stage          & 29.55         & 5.98     & 0.2319 \\
        \quad w/o align CLIP          & 33.78         & 4.37     & 0.2191 \\
        \quad w/o classifier-free          & 30.38         & 4.67     & 0.2359 \\
        \quad w/o loss decay      & 29.71         & 5.06     & 0.2257 \\
        \hline
    \end{tabular}
    \caption{\textbf{Ablation study of model architectures.} We compare the performance of full DiffDance and several architecture variants.}
    \label{tab:table2}
    \vspace{-2mm}
\end{table}

\noindent\textbf{Beat Align Score.} It is important for dance generation approaches to produce dance sequences that align well with input music. Our DiffDance achieves the best beat align score of $0.2418$, which outperforms $3.69\%$ over Bailando. We even obtain a better beat align score than ground truth ($0.2418$ \emph{v.s.} $0.2374$). As shown in Fig.~\ref{fig:beats_align}, the distance between music beats and dance beats of our Diffdance is smaller than ground-truth beats. This reflects that our model can generate dance sequences that tightly follow the beats of music, and has a deep understanding of music semantics corresponding to dance movements, indicating the effectiveness of aligning the Wav2CLIP music embedding space to motion and the use of classifier-free guidance.

However, it is crucial to note that this does not necessarily imply that DiffDance surpasses human performances in every aspect of dance. The BAS strictly rewards dance beats synchronized with music beats, yet a precise one-to-one mapping between the two doesn't invariably exist. Human performances may excel in other aspects of dance, such as expressiveness, creativity, and emotion, without necessarily maintaining a perfect alignment with the musical beat. This observation suggests that a new metric is needed in the future for further assessment of rhythmic alignment between music and dance.

\noindent\textbf{User Study.} Compared with objective evaluations for dance generation, subjective evaluation can provide a comprehensive performance comparison. Therefore, we conduct extensive user studies where we ask the participants to choose their preferred dance sequence generated by each previous method (including ground truth) and our DiffDance. Specifically, we invite $30$ participants with diverse demographic backgrounds, including $10$ experts with expert knowledge in dance choreography and $20$ non-experts. Using the AIST++ test set, we randomly crop $100$ music clips of $20$ seconds and generate $100$ music-dance pairs for each method, resulting in $700$ pairs in total. For each compared method, each participant is randomly assigned $10$ dance sequences together with corresponding dance sequences generated by our DiffDance with the same input music. We ask the participant to indicate the preferred one by ``\emph{Considering the music genres, rhythm, motion harmony, and diversity, which dance sequence overall is better aligned with the music}'' in every two candidates. Table~\ref{tab:tabel1} shows the statistic results, where our DiffDance outperforms all other methods with at least $74.4\%$ winning rate. It is noteworthy that many participants find that dance sequences generated by our method usually have distinct long-term choreographic structures, such as repeating the same action in different directions.

\begin{table}
    \centering
    \begin{tabular}{lccc}
        \hline
        Method  & FID$_k$$\downarrow$ & Div$_k$$\uparrow$ & BAS$\uparrow$ \\
        \hline
        MDM-base          & 54.35         & 3.32     & 0.2266 \\
        \quad + position          & 41.55         & 3.87     & 0.2631 \\
        \quad + position velocity         & 24.87         & 5.55     & 0.2412 \\
        \quad + rotation           & 35.88         & 6.89     & 0.2257 \\
        \quad + rotation velocity          & 28.50         & 5.41     & \textbf{0.2636} \\
        \quad + ALL     & \textbf{19.47}         & \textbf{7.09}     & 0.2384 \\
        \hline
    \end{tabular}
    \caption{\textbf{Ablation study of loss functions.} We compare the performance of our base music-to-dance model trained with different losses.}
    \label{tab:table3}
    \vspace{-2mm}
\end{table}

\begin{table}
    \begin{center}
    \begin{tabular}{cccc}
        \hline
        Method  & FID$_k$$\downarrow$ & Div$_k$$\uparrow$ & BAS$\uparrow$ \\
        \hline
        Ground Truth     & 17.10     & 8.19  & 0.2374  \\
        \hline
        $s=0$          & 24.67         & 5.88     & 0.2145 \\
        $s=10$          & 24.35         & 5.81     & 0.2223 \\
        $s=20$          & 24.38         & 5.88     & 0.2309 \\
        $s=30$    & \textbf{24.09}         & \textbf{6.02}     & \textbf{0.2418} \\
        $s=40$    & 24.57         & 5.80     & 0.2327 \\
        \hline
    \end{tabular}
      \end{center}
    \caption{\textbf{Ablation study of conditioning augmentation.} We compare various diffusion timesteps $s$ used in the SSR model during sampling. $s=0$ represents the model without conditioning augmentation.}
    \vspace{-2mm}
    \label{tab:table4}
\end{table}

\begin{figure*}[t]
\begin{center}
    \includegraphics[width=\linewidth]{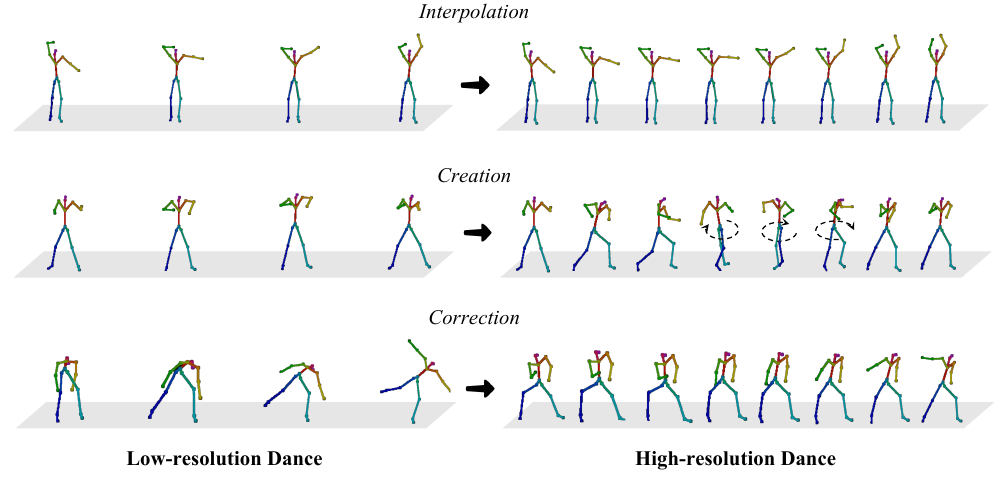}
    \caption{Visualization of outputs from both stages of our cascaded model. The base M2D model generates low-resolution dance sequences containing key motions. The SSR model improves low-resolution outputs by interpolating dance frames for increased temporal resolution, producing novel and meaningful relative dance frames, and rectifying ataxic poses. This highlights the advantages of our two-stage configuration.}
    \label{fig:amssr}
    \vspace{-2mm}
\end{center}
\end{figure*}

\subsection{Ablation Studies}


\noindent\textbf{Model Architecture.} 
We explore the effectiveness of the following $4$ architecture settings as shown in Table~\ref{tab:table2}. 
a) Two-stage pipeline. Compared to the one-stage model as `w/o two-stage', DiffDance improves FID$_k$ by $5.46$ ($22.67\%$), which reflects that the cascaded framework is essential for high-resolution dance generation. 
We also present the qualitative results of both stages to demonstrate how our cascaded framework enhances the overall performance of dance generation. As depicted in Fig.~\ref{fig:amssr}, we visualize three distinct functions of the SSR model. First, the SSR model can increase the temporal resolution by generating meaningful dance frames as interpolations, which is its primary function. Second, the SSR model can create new relative motion inspired by low-resolution dance frames, improving the motion diversity of DiffDance. Third, the SSR model can further correct unreasonable and uncoordinated motion frames generated by the low-resolution model, enhancing the robustness of DiffDance. By integrating these functions, our cascaded framework is able to generate realistic dance movements.
b) Alignment of music and motion embedding spaces. Compared to `w/o align CLIP', DiffDance improves FID$_k$ by $9.69$ ($40.22\%$). This indicates the effectiveness of aligning music and motion embedding spaces for better semantic music representation before training DiffDance. 
c) Classifierr-free guidance. Compared to `w/o classifier-free', DiffDance improves FID$_k$ by $6.29$ ($26.11\%$). 
d) Dynamic loss decay weight. Compared to `w/o loss decay', DiffDance improves FID$_k$ by $5.62$ ($23.33\%$).

\noindent\textbf{Loss Function.}
We conduct loss-setting experiments on the first stage base M2D model and evaluate the model on the low-resolution ground truths for efficient comparisons. The results for the whole cascaded pipeline can be extended similarly. We demonstrate the results in Table~\ref{tab:table3}. For only using losses in MDM~\cite{mdm} denoted as `MDM-base', FID$_k$ is only $54.35$, and fluctuates dramatically during the whole training process, producing highly jittery dance sequences. With regularizing the key joints in Equation~\ref{key_pos} and~\ref{key_rot} and adding `position', `position velocity', `rotation' and `rotation velocity' losses separately, FID$_k$ improves to $41.55$, $24.87$, $35.88$ and $28.50$ respectively. Finally, by adding all these geometric losses, we achieve the best FID$_k$ of $19.47$.

\noindent\textbf{Conditioning Augmentation.}
For conditioning augmentation, we compare various diffusion timestep $s$ to corrupt dance generated by the base M2D model as input condition of the SSR model. We show the results in Table~\ref{tab:table4}, where we sweep the timestep $s$ from $0$ to $40$. Note that $s=0$ means no conditioning augmentation is used for training and testing. Our model achieves the best quality and diversity at $s=30$, which is $30\%$ of the SSR diffusion timestep $T$. This indicates that adding moderate amounts of noise augmentation is beneficial for the cascaded generation pipeline. We fixed this noise ratio during sampling for all other experiments.

\section{Conclusion}

In this paper, we introduce a cascaded motion diffusion framework called DiffDance for music-driven dance generation. DiffDance comprises a base music-to-dance diffusion model and a sequence super-resolution diffusion model capable of generating high-resolution, long-form dance sequences with temporal consistency. To enhance semantic music representation, we align the music embedding space with motion by fine-tuning our music encoder using a contrastive objective. Additionally, we employ classifier-free guidance in the music-to-dance diffusion process. We also incorporate various geometric losses and a dynamic loss decay weight to improve the fidelity and diversity of dance samples. Comprehensive experimental results demonstrate the superiority of our method from both qualitative and quantitative perspectives. 

\section*{Acknowledgement}
This work was supported in part by the National Key R\&D Program of China under Grant 2022ZD0115502, the National Natural Science Foundation of China under Grant 62122010, and the CCF-DiDi GAIA Collaborative Research Funds for Young Scholars.

\bibliographystyle{ACM-Reference-Format}
\bibliography{MM--23}

\end{document}